\documentclass[12pt]{iopart}
\usepackage{graphicx} 
\usepackage{dcolumn} 
\usepackage{bm}
\newcommand{\ignore}[1]{} 
\usepackage{epsfig} 
\usepackage{color}
\usepackage{iopams}
\usepackage{multirow}

\begin{document}

\title[Opinion dynamics in two dimensions]{Opinion dynamics in two 
dimensions: domain coarsening leads to stable bi-polarization and anomalous 
scaling exponents} 

\author{F. Vel\'asquez-Rojas and F. Vazquez}

\address{IFLYSIB, Instituto de F\'isica de
  L\'iquidos y Sistemas Biol\'ogicos (UNLP-CONICET), 1900 La Plata,
  Argentina}

\ead{fede.vazmin@gmail.com}

\begin{abstract} 
We study an opinion dynamics model that explores the
competition between persuasion and compromise in a population of agents with
nearest-neighbor interactions on a two-dimensional square lattice.  Each agent
can hold either a positive or a negative opinion orientation, and can have two 
levels of intensity  --moderate
and extremist.  When two interacting agents have the same orientation become
extremists with persuasion probability $p$, while if they have opposite
orientations become moderate with compromise probability $q$.  These updating
rules lead to the formation of same-opinion domains with a coarsening dynamics
that depends on the ratio $r=p/q$.  The population initially evolves to a
centralized state for small $r$, where domains are composed by moderate agents
and coarsening is without surface tension, and to a bi-polarized state for
large $r$, where domains are formed by extremist agents and coarsening is
driven by  curvature.  Consensus in an extreme opinion is finally reached in a
time that scales with the population size $N$ and $r$ as 
$\tau \simeq r^{-1} \ln N$ for small $r$ and as $\tau \sim r^2 N^{1.64}$ for
large $r$.  Bi-polarization could be
quite stable when the system falls into a striped state where agents organize
into single-opinion horizontal, vertical or diagonal bands.  An analysis of
the stripes dynamics towards consensus allows to obtain an approximate 
expression for $\tau$ which shows that the exponent $1.64$ is a result of the
diffusion of the stripe interfaces combined with their roughness properties.   
\end{abstract}

\maketitle

\section{Introduction}
\label{intro}

In 1964, Abelson \cite{Abelson-1964} used a mathematical model to pose a
puzzle that still intrigues theoretical social scientists. He demonstrated
that convergence on ``monoculture'', an overall opinion consensus at the
population level, is inevitable in a connected population of individuals that
continuously update their views by moving towards the average opinion of their
neighbors.  However, extensive research on opinion formation shows that most
empirical opinion patterns resembles those of bi-polarization, rather than
those of consensus \cite{Abelson-1964}.   The phenomenon of  bi-polarization
is defined as the development of two groups with antagonistic  opinions that
intensify their differences over time, and where positions between the two
extremes of the opinion spectrum are increasingly less occupied (see
\cite{Mas-2013-2} for a recent review). The theoretical inevitability of
consensus, poorly supported by empirical  observations, lead Abelson to wonder:
``What on earth one must assume in order to generate the bimodal outcome of
community cleavage studies?''.  This is one of the long standing questions in
theoretical sociology.  In the same line, Bonacich and Lu \cite{Bonacich-2012}
have recently noted  that many models show how groups arrive to consensus, but
there are not generally accepted models of how groups become polarized or how
two groups can become more and more different and possible hostile.    Some
models that combine positive and negative social influence
\cite{Mark-2003,Flache-2011,VazMartins-2010} lead to a bimodal opinion
distribution that could explain bi-polarization.  However, negative  influence
is not fully supported by empirical evidence.

Based on previous works \cite{Lau-1998}, M\"as and Flache have recently
proposed in Refs. \cite{Mas-2013-2,Mas-2013-1} an alternative mechanism that
combines homophily \cite{McPherson-2001,Ibarra-1992} with 
``persuasion argument theory'' (PAT)
\cite{Isenberg-1986,Vinokur-1978,Myers-1982}, which gives rise to
bi-polarization without the assumption of negative influence.  The authors have
also performed group-discussion experiments to test the validity of the
theoretical model.  The idea is that, due to homophily an individual tends to
interact and talk with a partner that holds the same opinion  orientation on a
given issue, as for instance to be in favor of the same-sex  marriage.  Then,
PAT suggests that the two interacting individuals are likely to exchange
different arguments that support their positions, and thus they can provide
each other with new arguments or reasons which reinforce their initial
opinions.  This could intensify the individuals' views and make them more
extreme in their believes.  Motivated by this work, La Rocca et al.
\cite{LaRocca-2014} have recently introduced a model that incorporates the
mechanisms of homophily and persuasion in a simple way, and that is able to
generate desired levels of bi-polarization.  We refer to this model as the
``M-model'' from now on.   The opinion of each agent is represented by an
integer number $k$ bounded in the interval $[-M,M]$ ($k \neq 0$) that
describes its degree of agreement on a political issue, from totally against
($k=-M$) to totally in favor ($k=M$).  Each agent is allowed to interact with
any other agent in the  population, which corresponds to a mean-field (MF)
setup (all-to-all interactions).  Two interacting agents with the same
orientation (positive or negative) reinforce their opinions in one unit and
become more extremists with persuasion probability $p$, while the opinions of
two interacting agents with opposite orientations get two units closer with
compromise probability $q$.  It was shown in  \cite{LaRocca-2014} that the
behavior of the model depends on the relative frequency between
same-orientation (persuasion) and opposite-orientation (compromise)
interactions, determined by the ratio $p/q$.  When persuasive events are more
frequent than compromise events, opinions are driven towards extreme values
$k=-M$ and $k=M$, inducing the coexistence of extreme opinions or
bi-polarization.  In the opposite case, when compromise events dominate over
persuasion events, opinions are grouped around moderate values $k=-1$ and
$k=1$, leading to centralization.  Also, it was observed that stationary
states of bi-polarization and centralization are unstable, given that a small
opinion asymmetry is enough to drive the  population to a fast consensus in
one of the two extreme opinions.  While these results correspond to the MF
version of the M-model, the consequences of the competition between persuasion
and compromise have not been explored in spatial or complex interaction
topologies.

In this article we study the dynamics of the M-model on a two-dimensional
($2D$) square lattice, for the simplest and non-trivial case $M=2$.  Our goal
is to  investigate the effects of the persuasion and compromise mechanisms in
a population of agents with nearest-neighbor (short range) interactions, in
contrast to the all-to-all interactions of the MF case.  In particular, we aim
to explore how the $2D$ spatial topology affects the stability of the
polarized and centralized states.  We also aim to understand basic properties
of the approach to extremist consensus.

The mechanisms of persuasion and compromise have been implemented in several
works to model opinion formation in interacting populations.  On the one hand,
persuasion have recently been introduced in some agent-based models
\cite{Crokidakis-2012,Crokidakis-2013,Balenzuela-2015,Terranova-2014}.  For
instance, persuasion was used in Refs. \cite{Crokidakis-2012,Crokidakis-2013}
as a degree of a person's
self-conviction where, in addition to the influence from others, a person
takes into account its own opinion when making a decision.  The authors in
Ref. \cite{Balenzuela-2015} introduced a model where each individual can have
one of two opposite opinions or be undecided, and each of these three choices
is determined by its persuasion or degree of conviction on the given issue,
represented by a real number on a persuasion interval.  Another work studied a
model where the persuasion takes place between opposite-orientation agents
\cite{Terranova-2014}.  On the other hand, the compromise process was
initially studied in models with continuous opinions and interaction
thresholds \cite{Weisbuch-2002,Ben-Naim-2003}, and the stability of the
bimodal opinion distribution was tested under the influence of noise
\cite{Pineda-2009}.  Some multistate voter models
\cite{Castello-2006,Volovik-2012,DallAsta-2008,Vazquez-2008} have incorporated
a rule similar to compromise that uses a reinforcement mechanism by which
agents switch orientation only after receiving multiple inputs of agents with
opposite orientation.  For instance,  Castell\'o et al. \cite{Castello-2006}
studied a three-state language model where each agent could either speak one
of two possible languages (A or B) or be a bilingual AB.  A monolingual
A-agent can become bilingual AB by interacting with an agent that speaks the
opposite language (B-agent or AB-agent).  They  investigated the ordering
process of the system  on regular lattices and small world networks, and the
mean consensus  time associated to each topology.  A stability analysis of
this model  \cite{Vazquez-2010} revealed that the dominance of one language is
enhanced by the connectivity of the network, and that this effect is even
stronger in lattices.  More recently, Volovik and Redner \cite{Volovik-2012}
studied a voter model with four states, in which each agent can choose between
two possible opinions and can additionally have two levels of commitment to the
opinion  (confident and unsure).  A confident voter that interacts with an
agent of a  different opinion becomes less committed (unsure), but keeps its
opinion.  However, an unsure voter can change its opinion by interacting with
an agent of a different opinion.  In another work \cite{DallAsta-2008}
Dall'Asta and Galla performed a numerical and analytical study of the
coarsening properties of general voter models with many intermediate states on
lattices \cite{Vazquez-2008}, which have interaction rules similar to that of
the works \cite{Castello-2006,Volovik-2012} described above.  They showed that
the addition of intermediate states to the 2-state voter model (VM)
\cite{Liggett-1975} restores an effective surface tension.  It is important to
mention that all these models lack the mechanism of  strengthening of opinions
induced by the same-orientation interactions that characterizes the M-model.

The rest of the paper is organized as follows.  In section~\ref{model} we
describe the M-model on a $2D$ square lattice.  In section~\ref{coarsening} we
analyze the temporal evolution of the system and explore the coarsening
dynamics in the regimes of bi-polarization and centralization.  Results on the
behavior of the mean consensus times are presented in section~\ref{consensus}.
In section~\ref{stripes} we investigate the dynamics of interfaces between
opinion domains in the large persuasion limit.  This study allows to derive an
approximate expression for dependence of the mean consensus time with the
system size, which explains the non-trivial scaling observed in the M-model
and in general models with coarsening by surface tension.  Finally, in
section~\ref{summary} we summarize and discuss our  findings.

\section{The M-model on a square lattice}
\label{model}

We consider the opinion formation dynamics of the model proposed by La Rocca
et al. \cite{LaRocca-2014} on a $2D$ square lattice of $N=L^2$ sites, where $L$
is the linear size of the lattice.
Each site is occupied by an agent who can interact with its four
nearest neighbors, and can take one of four possible opinion states
$k=-2,-1,1$ or $2$ that represents its position on a political issue, from a
negative extreme $k=-2$ (a negative extremist) to a positive extreme $k=2$ (a
positive extremist), taking moderate values $k=-1,1$ (a moderate).  The sign
of $k$ and its absolute value $|k|$ indicate the opinion orientation and its
intensity, respectively.  In a single time step of the dynamics of length
$\Delta t=2/N$, two nearest-neighbor agents $i$ and $j$ with respective states
$k_i$ and $k_j$ are picked at random to interact.  Then, their states are
updated according to their opinion orientations (see Fig.~\ref{pers-comp}):
 
\begin{figure}[h]
  \begin{center}
    \centerline{\includegraphics[width=14.5cm]{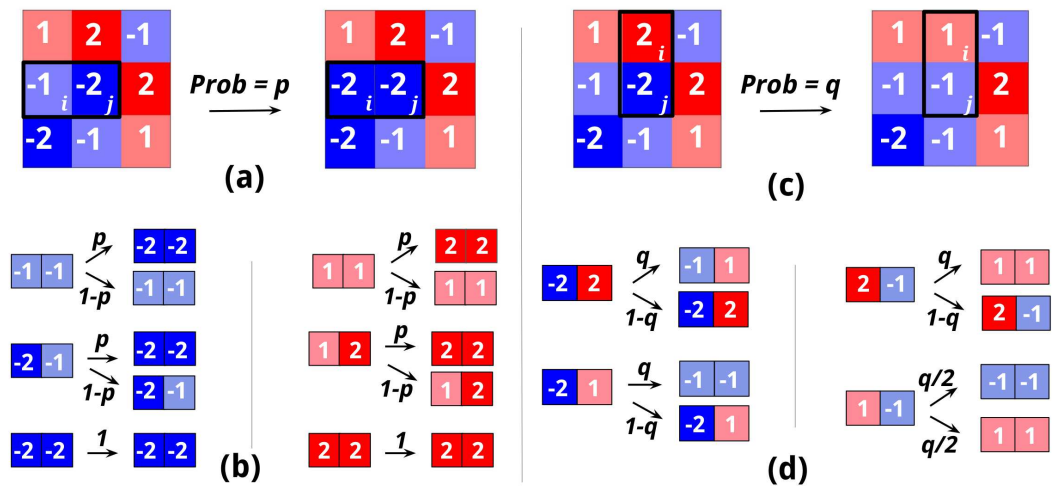}}
    \caption{The two update events of the M-model on a square lattice.  (a)
      and (b) Persuasion.  In panel (a), a positive moderate agent $i$ that has
      opinion $k_i=-1$ becomes extremist ($k_i=-1 \to k_i=-2$) with persuasion
      probability $p$ by interacting with a nearest-neighbor agent $j$ that has
      extreme opinion $k_j=-2$.  Panel (b) shows all possible persuasion events in
      which two neighboring  agents with the same opinion orientation reinforce
      their opinions and become extremists.  (c) and (d) Compromise.  In panel (c),
      two interacting neighbors with opposite and extreme opinions become moderate
      with probability $q$ ($k_i=2 \to k_i=1$ and $k_j=-2 \to k_j=-1$).  Panel (d)
      shows all possible compromise events  in which two neighbors with opposite
      orientations become moderate.}
    \label{pers-comp}
  \end{center}
\end{figure}

\begin{itemize}
\item \emph{Persuasion} [Figs.~\ref{pers-comp}(a) and \ref{pers-comp}(b)]:  If
they have the same orientation  ($k_i,k_j>0$ or $k_i,k_j<0$), then a
persuasion event happens with probability $p$.  An agent increases its
intensity by one if it is a moderate ($|k|=1$), while it keeps its opinion if
it is an extremist ($|k|=2$).
 \item \textit{Compromise} [Figs.~\ref{pers-comp}(c) and \ref{pers-comp}(d)]:
If they have opposite orientations  ($k_i>0$ and $k_j<0$ or $k_i<0$ and
$k_j>0$), then a compromise event happens with probability $q$.  If both
agents are extremists ($|k_i|=|k_j|=2$) they decrease their intensities by
one.  If one is an extremist and the other is a moderate $|k|=1$, then the
extremist decreases its intensity by one while the moderate switches
orientation.  If both agents are moderates one switches orientation at random.
\end{itemize}

We can think of persuasion and compromise as two competing mechanisms that
shape the distribution of opinions in the population.  While  persuasive
interactions make individuals adopt extreme opinions $2$ and $-2$ and lead to
opinion bi-polarization, compromise contacts tend to moderate opinions, promoting
a centralized opinion  distribution around moderate values $1$ and $-1$.

\section{Coarsening dynamics}
\label{coarsening}

\begin{figure}[t]
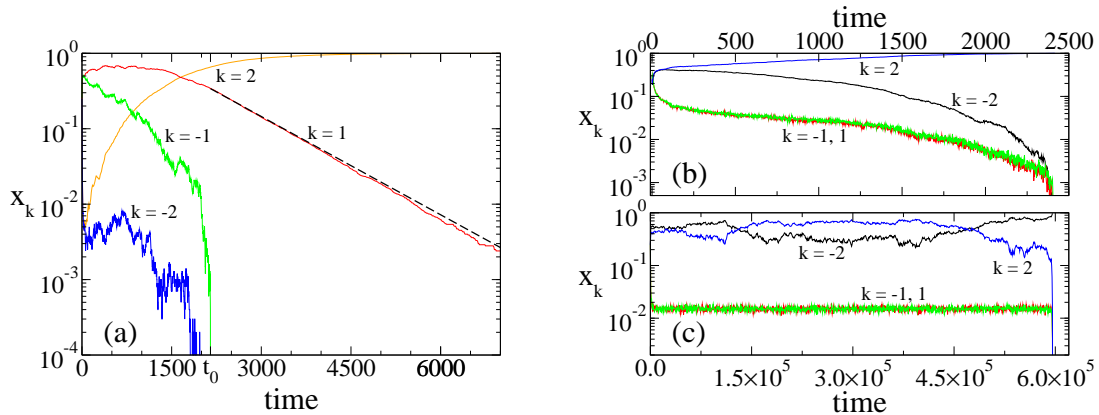

  \begin{center}
    \begin{tabular}{cc}
      \includegraphics[width=4.5cm, bb=70 -20 550 550]{Fig2a.eps} &
      \hspace{2.5cm}
      \includegraphics[width=4.5cm, bb=70 -20 550 550]{Fig2bc.eps}
    \end{tabular}
    \caption{Time evolution of the densities $x_k$ of agents in different
      opinion states $k$ in single realizations of the dynamics, for a system of
      $N=10^4$ agents and two values of $r=p/q$.  (a) $x_k(t)$ for $r=10^{-3}$.  The
      dashed line is the expression $0.34 \, e^{-r(t-2150)}$ from
      Eq.~(\ref{x1-t}).  Panels (b) and (c) show $x_k(t)$ for $r=1/3$ in   
      realizations of type 1 and type 2, respectively.}
    \label{nk-t}
  \end{center}
\end{figure}

We started the analysis of the model by studying the time evolution of the
number of agents in each state, which describes the system at the macroscopic
level.  For that, we run Monte Carlo  simulations of the dynamics described in
section \ref{model} and measured the quantities $x_k(t)$ ($k=-2,-1,1,2$),
defined as the fraction of agents in state $k$ at time $t$, which are
normalized at all times ($\sum_k x_k(t)=1$ for all $t\ge 0$).  Initially, 
each agent adopts one of
the four possible states with equal probability $1/4$.  The qualitative
behavior of the system turns out to depend on the relative frequency between
persuasion and compromise events, which is controlled by the ratio $r \equiv
p/q$ between persuasion and compromise probabilities.  Therefore, for
convenience we set $p+q=1.0$ [$p=r/(1+r)$ and $q=1/(1+r)$] and analyzed the
system as $r$ is varied.    In Fig.~\ref{nk-t} we show the evolution of the
densities $x_k$ in single realizations of the dynamics for a system of size
$N=10^4$, and two different values of $r$.

In the realization with a  very small $r=10^{-3}$ [Fig.~\ref{nk-t}(a)]
compromise interactions are much more frequent than persuasive interactions
($q \gg p$), driving most agents' opinions  towards moderate values during an
initial stage ($t \lesssim 500$) in which $x_1$ and $x_{-1}$ are much larger
than $x_2$ and $x_{-2}$.  This corresponds to a centralization of opinions.
Then, at time $t_0=2150$ negative states $-1$  and $-2$ dissappear and the
density $x_{1}$ decays exponentially fast to zero, while
$x_{2}$ approaches exponentially fast to $1$.  Once $x_{2}$ equals $1$ the
system cannot longer evolve (absorbing state), which in this case corresponds
to a positive extremist opinion consensus.  In a general case, the ultimate
state of the system is always consensus in either extremist state, i e., all
agents with opinion  $2$ ($x_2=1$) or all with opinion $-2$ ($x_{-2}=1$).
An insight into this exponential approach to consensus can be obtained within
a mean-field (MF) approximation which assumes that every agent interacts with
every other  agent.  This corresponds to the MF version of the M-model for
small $r$ studied in \cite{LaRocca-2014}.  After time $t_0$ only positive
states $1$ and $2$ remain in the system ($x_1(t)+x_2(t)=1$), and thus the
dynamics is only driven by persuasive events that slowly drive all agents to
state $2$  with a very small probability $p=r/(1+r) \simeq r$ in the $r \ll 1$
limit.  Then, the mean change of $x_1$ in a single time step of length
$\Delta t=2/N$ is given by
\begin{eqnarray} 
  \frac{dx_1}{dt} = \frac{\Delta x_1}{\Delta t} = -\frac{p \,
    x_1^2 \, \frac{2}{N}}{2/N} - \frac{p \, 2 \, x_1 \, x_2 \frac{1}{N}}{2/N} =
  -p \, x_1 ~~~~\mbox{for $t \ge t_0$}.
  \label{dx1dt}
\end{eqnarray}    
The first term of Eq.~(\ref{dx1dt}) describes the
interaction between two state-$1$ agents that make the transition to state $2$
with probability $p$, while the second term accounts for the transition to
state $2$ of a state-$1$ agent that interacts with a state-$2$ agent.  The
solution of Eq.~(\ref{dx1dt}) is
\begin{equation} 
  x_1(t) = x_1(t_0) \, e^{-r(t-t_0)} ~~~~\mbox{for $t \ge t_0$},
  \label{x1-t}
\end{equation} 
where we have used $r$ as an approximate value for $p$.
Expression Eq.~(\ref{x1-t}) for $x_1$ is plotted in Fig.~\ref{nk-t}(a) (dashed
line) using $r=10^{-3}$ and the initial condition $x_1(t_0=2150) \simeq 0.34$
extracted from the curve of $x_1(t)$.  The good agreement with simulations
shows that the dynamics on the lattice for small $r$ is well described by the
MF theory.

In the realizations with $r=1/3$ [Figs.~\ref{nk-t}(b) and \ref{nk-t}(c)] persuasive
interactions, which are more frequent than in the previous case but still less
often than compromise interactions, seem enough to make most agents adopt extreme
states $2$ and $-2$, and thus $x_{2}$ and $x_{-2}$ are larger than $x_{1}$ and
$x_{-1}$ for all times.  This corresponds to a polarized state where the
population of agents is divided in two groups of similar size that hold
extreme and opposite opinions.  We also observe that in the realization of
panel (b) the system reaches consensus in extremist state $2$ at time $t
\simeq 2400$, while in the realization of panel (c) an extremist consensus in
state $-2$ is achieved in a much longer time $t \simeq 6 \times 10^5$.  These
examples correspond to two different types of realizations observed in
simulations.  In  realizations of type $1$ [panel (b)] the initial symmetry
between positive and negative states is broken at early times and the system
is quickly driven towards consensus, where the densities $x_1$ and $x_{-1}$
decay to zero and either $x_2$ or $x_{-2}$ approaches $1$.  In realizations of
type $2$ [panel (c)] the system falls in a long-lived metastable state where
$x_1$ and $x_{-1}$ fluctuate around a stationary value for a very long time
until they drop to zero together with $x_2$.  This metastable state lasts for
a much longer time than the one observed in the MF version of the model 
\cite{LaRocca-2014}.  This means that the
coexistence of opinions could be very stable when interactions are restricted
to nearest-neighbors on a lattice, increasing the stability of the opinion 
bi-polarization.

\begin{figure}[t]
  \begin{center}
    \begin{tabular}{ccc} \hspace{0.5cm}
      \includegraphics[width=3.9cm, bb=70 -20 550 550]{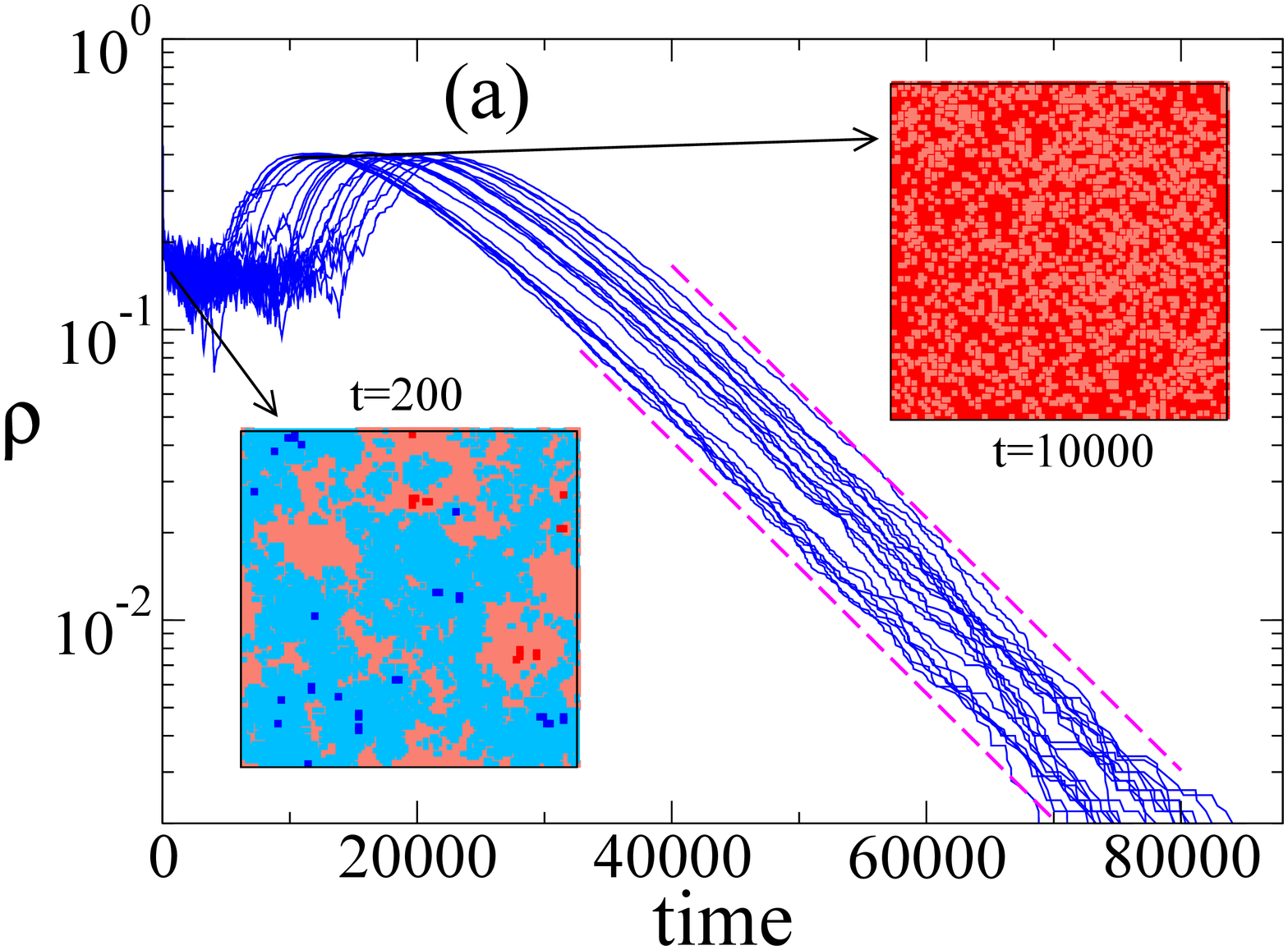} &
      \hspace{1.6cm}
      \includegraphics[width=3.9cm, bb=70 -20 550 550]{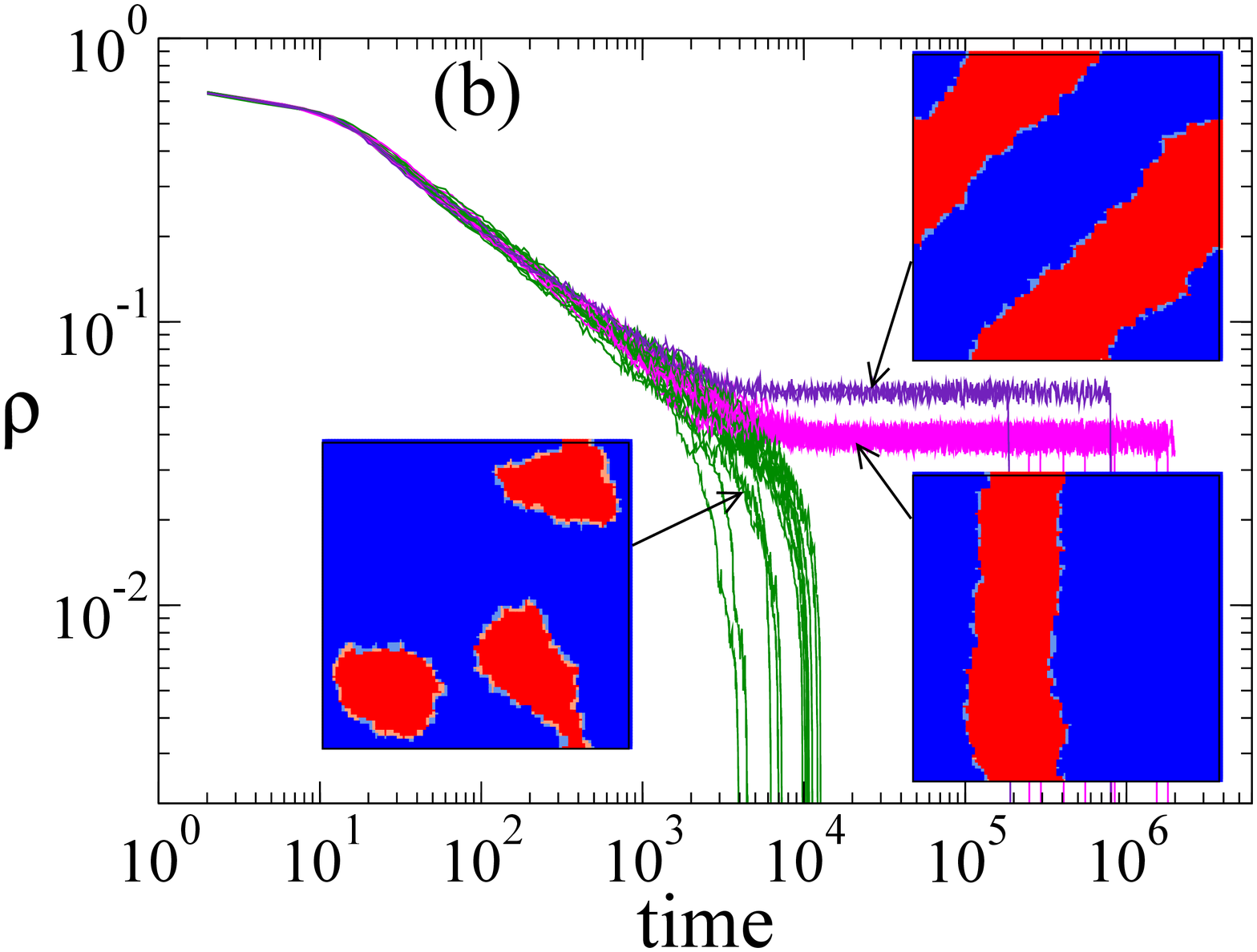}     &
      \hspace{1.6cm}
      \includegraphics[width=3.9cm, bb=70 -20 550 550]{Fig3c.eps}
    \end{tabular}
    \caption{Time evolution of the interface density $\rho$ in single
      realizations, for a system of size $N=10^4$ and $r=10^{-4}$ (a) and $r=1/3$
      (b).  The snapshots of the lattice show the spatial pattern of 
      opinions at different times and for different realization types.  Panel
      (a): the down-left
      snapshot corresponds to the centralization of opinions around moderate values
      $k=-1$ and $k=1$, while in the top-right snapshot all opinions are positive
      and driven by persuasion.  Dashed lines have slope $r=10^{-4}$.  Panel (b):
      the down-left snapshot corresponds to realizations that reach a quick
      consensus by domain coarsening (type 1), while down-right and top-right
      snapshots represent realizations of type 2, where the system gets trapped in a
      long-lasting stripe state before reaching consensus.  Panel (c): average
      interface density $\langle \rho \rangle$ (circles) and average density of
      moderate states (squares) vs time on a system of size $N=300^2$.  The average
      was done over $10^{4}$ realizations.  Dashed lines have slope $-0.46$.}
    \label{rho-t}
  \end{center}
\end{figure}

In order to investigate the origin of the different behaviors described above
we study the coarsening properties of the system by looking at the density of
interfaces $\rho$, defined as the density of bonds between neighbors in
different states \cite{Castello-2006,DallAsta-2008}.  In Fig.~\ref{rho-t} we
show the time evolution of $\rho$ in single realizations for $r=10^{-4}$
[panel (a)] and $r=1/3$ [panel (b)], together with snapshots of the lattice at
different times and for each type of realization.  We observe the formation of
same-opinion domains with different characteristics.  For $r=10^{-4}$
[Fig.~\ref{rho-t}(a)], the large frequency of compromise events as compared to
persuasive events drives almost all agents towards moderate states, leading to
the early formation of large domains composed by agents with states $1$ or
$-1$, with a few sparse extremists (down-left snapshot).  During this stage,
the dynamics at the
interface between $1$ and $-1$ domains follows that of the VM.  This
explains the noisy shape of the interface that characterizes the coarsening
without surface tension of the VM \cite{Dornic-2001}.  The domains
slowly growth in size until almost all agents --except for a few extremists-- 
adopt the same moderate state (state $1$ in the snapshot), and $\rho$ reaches
a minimum.  This corresponds to the beginning of the persuasive stage discussed
previously, during which moderate agents become extremists.  The final
relaxation to consensus follows the MF exponential decay $\rho \simeq
x_1(1-x_1) \sim e^{-r \, t}$ from Eq.~(\ref{x1-t}) (dashed lines).  We note
that this  dynamics is very different from that observed in related multistate
voter models \cite{Castello-2006,DallAsta-2008,Volovik-2012}, where agents
with intermediate (moderate) states place themselves at the boundaries between
extreme-state domains and form rather smooth interfaces.  This last phenomenon
happens for $r=1/3$  [Fig.~\ref{rho-t}(b)], where moderate states $1$ and $-1$
are located at the interface between $2$ and $-2$ domains.  This is checked in
Fig.~\ref{rho-t}(c) where we show the time evolution  of the average value of
$\rho$ and the average density of moderate (intermediate) states $x_1+x_{-1}$.
We see that both $\langle \rho \rangle$ and $\langle x_1 + x_{-1} \rangle$
decay as $t^{-0.46}$, indicating that the interface dynamics is correlated
with that of the moderate  states.  The behavior $\langle \rho \rangle \sim
t^{-0.46}$ is consistent with the algebraic coarsening found in voter models
with intermediate states \cite{Castello-2006,DallAsta-2008}.  As it was shown
in \cite{DallAsta-2008}, the addition of intermediate states to the 2-state
VM changes the phase-ordering properties of the system, from a
coarsening driven by interfacial noise observed in the VM to a coarsening
driven by surface tension in models with one or many intermediate states.
Also, the coarsening exponent $0.46$ is compatible with the exponent $0.5$
associated to the domain growth driven by curvature observed in kinetic Ising
models \cite{Gunton-1983,Bray-2002}.

Another observation from Fig.~\ref{rho-t}(b) is related to the different types
of realizations, whose interface dynamics explains the temporal behavior of
the moderate densities $x_1$ and $x_{-1}$ observed in Figs.~\ref{nk-t}(b) and 
\ref{nk-t}(c).  The initial evolution of $\rho$ in all realizations follows
the power-law decay described above, but then they split in two main groups.  
The group
of short-lived realizations corresponds to Fig.~\ref{nk-t}(b), in which small
domains shrink and dissappear until one large extremist domain covers the
entire lattice (down-left snapshot).  The group of realizations that fall into
a long-lived metastable state, which consists on either horizontal stripes or
vertical stripes (down-right snapshot) or diagonal stripes  (top-right
snapshot), corresponds to Fig.~\ref{nk-t}(c).   In these dynamical metastable
states, the interface density $\rho$ fluctuates around a stationary value
until a  finite-size fluctuation takes the system to one of the absorbing
states ($\rho=0$).  The long plateau observed in $\rho$ shows that the
polarized state is much more stable in lattices than in MF
\cite{LaRocca-2014}.  As we shall study in more detail in section \ref{stripes},
this behavior is due to the slow diffusion of the interfaces between these
stripes, which eventually meet and annihilate and lead the system to consensus.
Diagonal stripes are characterized by a stationary value of $\rho$ that is
approximately $\sqrt{2}$ times larger than the corresponding value for
horizontal or vertical stripes.  It is also worth mentioning that, even though 
diagonal stripes were not  reported in
related models \cite{Castello-2006,Volovik-2012,Chen-2005}  probably because
they are very unlikely to be formed (around $3$ percent of the time in our
simulations), we expect to see diagonal stripes in all these models with
Ising-like coarsening \cite{Cugliandolo-2016}.

\section{Consensus times}
\label{consensus}

\begin{figure} 
  \centerline{\includegraphics[width=7.5cm]{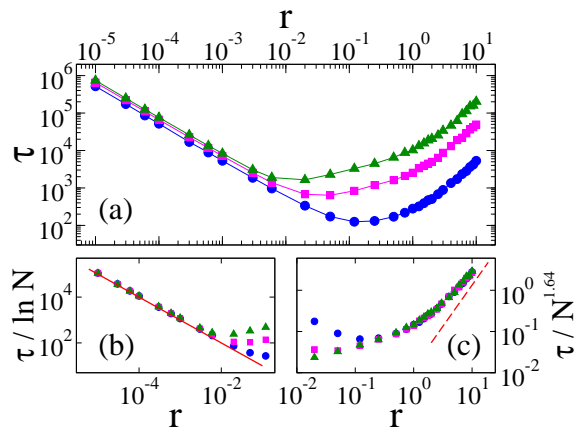}}
  \caption{(a) Mean consensus time $\tau$ as a function of $r$ on a double
    logarithmic scale for lattice sizes $N=100$ (circles), $N=400$ (squares)
    and $N=900$ (triangles).  Panels (b)
    and (c) show the data collapse for small and large $r$, respectively.
    The solid line in (b) is the analytical approximation $\tau \simeq r^{-1} \ln N$ 
    from Eq.~(\ref{tau-r-small}), while the dashed line in (c) has slope $2$.} 
  \label{tau-r}
\end{figure}

As we showed in section \ref{coarsening}, the M-model has two absorbing states
corresponding to the two extremist consensus.  A magnitude of interest in these
models is the mean time to reach opinion consensus $\tau$.  In
Fig.~\ref{tau-r}(a) we present results from numerical simulations of $\tau$ as
a function of $r$ and three different lattice sizes $N$.  Each data point
corresponds to an average over $10^4$ independent realizations with uniform
initial condition.  We see that $\tau$ has a non-monotonic shape with $r$,
taking very large values for small and large $r$.  We also observe that $\tau$
increases with $N$ and that the increase is much faster for large $r$, which
suggests two different scalings at both sides of the minimum.  Indeed, panels
(b) and (c) of Fig.~\ref{tau-r} show the collapse of the data at small and
large values of $r$ when curves are rescaled by $\ln N$ and $N^{1.64}$,
respectively.   The logarithmic scaling of $\tau$ with $N$ in the small $r$
limit can be obtained from the behavior of the density $x_1$ given by
Eq.~(\ref{x1-t}).  We first note that the exponential decay of $x_1$ with time
holds for any $r \ll 1$ and $N$ (not shown), and that the time $t_0(r,N)$ at
which the persuasive stage begins varies with both $r$ and $N$.  To derive an
expression for $\tau$ we make two assumptions.  First, we expect that the
distribution of states at $t_0(r,N)$ is peaked at $k=1$, i e., $x_1(t_0)
\simeq 1$ and $x_2(t_0) \simeq 0$.  Indeed, we have checked that $x_1(t_0)$
approaches $1.0$ as $r$ decreases ($x_1(t_0) \simeq 0.34$ for $r=10^{-3}$
while $x_1(t_0) \simeq 0.7$ for $r=10^{-4}$).   Second, we assume that the
consensus is reached when there is less than one agent in state $1$, which
leads to the condition $x_1=1/N$ at $\tau$.  Then, solving for $\tau$ from the
relation  $1/N \simeq e^{-r[t-t_0(r,N)]}$ we arrive to the approximation
$\tau \simeq t_0(r,N) + r^{-1} \ln N$.  The second term associated to the
duration of the persuasive stage dominates in the small $r$ limit and,
therefore, $\tau$ can be approximated as
\begin{equation} 
  \tau \simeq \frac{\ln N}{r} ~~~~ \mbox{for $r \ll 1$}.
  \label{tau-r-small}
\end{equation} 
We observe in Fig.~\ref{tau-r}(b) that the analytical
expression Eq.~(\ref{tau-r-small}) represented by a solid line has a good
agreement with numerical data, showing the $1/r$ divergence of $\tau$ in the
$r \to 0$ limit.

\begin{figure} 
  \centerline{\includegraphics[width=7.5cm]{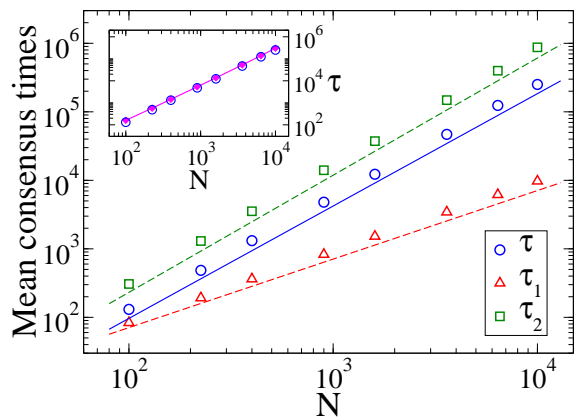}}
  \caption{Mean consensus times $\tau$, $\tau_1$ and $\tau_2$ vs system size
    $N$ on a log-log scale for $r=1/3$.  $\tau$ is the average over all $10^4$
    realizations, while $\tau_1$ and $\tau_2$
    correspond to the average values over realizations of type $1$ and type
    $2$, respectively.  Straight lines have slopes $1.71$, $1.64$ and $1.0$
    (from top to bottom).  Inset: the estimation of the mean consensus time as
    the linear combination $0.66 \, \tau_1 + 0.34 \, \tau_2$ (diamonds) of both
    realization types is compared to $\tau$ (circles).}
  \label{tau-N}
\end{figure}

The power-law behavior $\tau \sim N^{1.64}$ used to collapse the data for
large $r$ [see Fig.~\ref{tau-r}(c)] was obtained by running simulations for
$r=1/3$ and various system sizes.  Results are shown in Fig.~\ref{tau-N} with
empty circles, where we also plot a solid line with slope $1.64$ which serves
as a guide to the eye, corresponding to the best fit of the data.  The data
points of  Fig.~\ref{tau-r}(c) collapse into a single curve that seems to
approach  the quadratic behavior $r^2$ as $r$ becomes large (dashed line),
which surprisingly agrees with that predicted by the MF expression
$\tau_{\mbox{\tiny MF}} \sim r^2 \ln N$ derived in \cite{LaRocca-2014}.
However, this logarithmic increase  of $\tau_{\mbox{\tiny MF}}$ with $N$ in MF
is much slower than the non-linear increase $\tau \sim N^{1.64}$ obtained
in lattices.  As consequence, the consensus in MF is much faster than in
lattices.

\begin{figure} \centerline{\includegraphics[width=7.5cm]{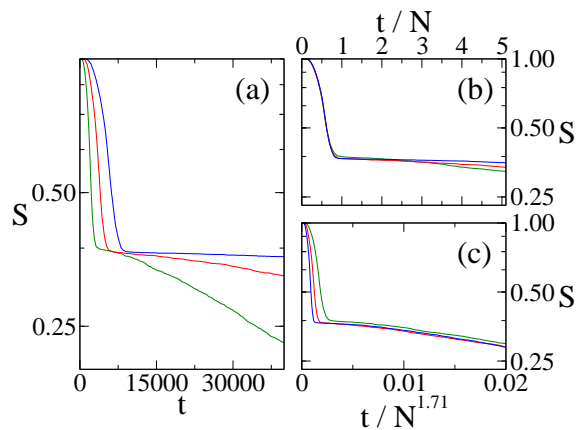}}
  \caption{(a) Survival probability $S$ vs time on a linear-log scale, for
    $r=1/3$ and system sizes $N=3600$, $6400$ and $10000$ (from bottom to top).  
    The initial fast decay of
    $S$ describes the domain coarsening, which has a mean lifetime
    proportional to $N$ [panel (b)].  The long exponential tail of $S$
    decays with a time constant proportional to $N^{1.71}$, associated to the
    mean lifetime of type-2 realizations [panel (c)].}
  \label{S-t}
\end{figure}

As we explain below, long consensus times in lattices for large $r$  are a
consequence of the long-lived metastable states that characterize the
realizations of type 2 discussed in  section \ref{coarsening}, which lead to
the non-trivial scaling exponent $1.64$.  Indeed, the value of $\tau$ obtained
from simulations is the combination of two main types of
realizations that have very different time scales.  That are, realizations of
type 1 where consensus is reached by domain coarsening, and realizations of
type 2 in which  consensus is reached by the diffusion of the two interfaces
that define the stripe.   To distinguish between type-1 and type-2
realizations we follow the method developed in \cite{Chen-2005,Castello-2006}
and study the distribution of consensus times $P(t)$, from where the mean
consensus time is calculated as  $\tau = \int_0^{\infty} t\, P(t) \, dt$.
This is equivalent to study the survival probability $S(t)$ of single runs
defined as the probability that a realization did not reach consensus up to
time $t$, which is related to $P(t)$ by the expression $S(t)=1- \int_0^t P(t)
\, dt$.  The advantage of calculating $S$ instead of $P$ is that $S$ has less
fluctuations associated to the finite number of realizations.
Figure~\ref{S-t}(a) shows $S$ vs time for $r=1/3$ and three system sizes. In
agreement with results in related models \cite{Chen-2005,Castello-2006},
curves are characterized by two time scales --a short time scale consistent
with a fast decay to consensus, and a much longer time scale associated with
an asymptotic exponential decay (the tail).  The initial fast decay of $S$
corresponds to the consensus induced by   coarsening observed in type-1
realizations, while the exponential tail describes the consensus times of
realizations that get trapped in a stripe metastable state (type-2
realizations).  Then, the time $t^*$ at which the exponential decay begins was
taken as a reference to assign a type to a given realization.  Realizations
that reached consensus before (after) $t^*$ were considered of type 1
(type 2).  Using this criteria we estimated the time to reach consensus in
each type of realization.  In Fig.~\ref{tau-N} we show that the mean consensus
time scales as $\tau_1 \sim N$ in type-1 realizations, while the scaling
$\tau_2 \sim N^{1.71}$ was found for type-2 realizations.  The data collapse
in panels (b) and (c) of Fig.~\ref{S-t} shows that $\tau_1$ can be considered
as the characteristic time scale associated to the fast initial decay of $S$,
and that $\tau_2$ is proportional to the time constant of the exponential decay.
We have also calculated the probability that a realization reaches the
metastable state as the fraction of type-2 realizations over $10^3$
independent runs, which gave the approximate mean value $0.34$ in the size
range  $400 \le N \le 10000 $, with a very slow decrease as $N$ increases.
The indirect estimation of the mean consensus time as the combination of the
two realization types
\begin{equation} 
  \tau \simeq 0.66 \, \tau_1 + 0.34 \, \tau_2
  \label{tau}
\end{equation} 
is plotted in the inset of Fig.~\ref{tau-N} (solid  diamonds),
where we observe a good agreement with the value of $\tau$ calculated over all
realizations (empty circles).   Therefore, the approximate scaling $\tau \sim
N^{1.64}$ observed in simulations can be explained as the result of the linear
combination of the power-law behaviors $\tau_1 \sim N$ and $\tau_2 \sim
N^{1.71}$.  Since $\tau_2$ becomes much larger than $\tau_1$ as $N$ increases
--by a factor of $10$ ($100$) for $N=400$ ($10^4$), we expect that the effective
exponent $1.64$ of $\tau$  will approach the exponent of $\tau_2$ as $N$
increases.  In the next section we provide an explanation of the
non-trivial exponent $1.71$ by studying the dynamics of stripes in detail.

\section{The dynamics of stripes towards consensus}
\label{stripes}

In section \ref{consensus} we showed that the mean consensus time for $r=1/3$
scales as $\tau \sim N^{1.64}$ with the system size $N$.  As discussed
previously, this scaling is mainly due to the existence of metastable states
that survive  for very long times, in which the system exhibits a stripe-like
pattern.  It is important to mention that very similar scaling laws for the
consensus time with system size, $\tau \sim N^{\nu}$, were already reported in
the literature in related works in lattices
\cite{Chen-2005,Castello-2006,Volovik-2012}.  For instance, in the  Majority
Rule (MR) model introduced in \cite{Chen-2005} the authors found $\nu=1.7$,
while in the bilinguals model studied in \cite{Castello-2006} an exponent
$\nu=1.8$ was obtained, and also a similar exponent was observed in the
confident VM investigated in \cite{Volovik-2012}, whose exact value
was not reported.  What all these models have in common with the M-model on a
lattice is
the existence of stripe states with a probability around $1/3$ when the system
starts from random initial conditions, and an ultimate consensus state that is
absorbing.  Despite that these models differ in the number of opinion states
($2$ states in MR model, $3$ states in the bilinguals model, $4$ states in
the confident VM, and $4$ or more states in the M-model), their
microscopic rules induce a coarsening dynamics that is driven by surface
tension, which can lead to the formation of horizontal, vertical or diagonal
stripes in square lattices, as it is  known to happen in Ising-like systems
\cite{Cugliandolo-2016}.  Therefore, it seems that the dynamics of stripes is
the fundamental mechanism that determines the consensus times in lattice
models with coarsening by surface tension and frozen consensus states, leading
to the scaling $\tau \sim N^{\nu}$ (with $1.64 \le \nu \le 1.8$) reported in
the works mentioned above.  As far as we know, there is no yet a satisfying
explanation for the behavior of  $\tau$ with $N$.  Some attempts to obtain the
exponent $\nu$ were developed in \cite{Chen-2005} and \cite{Volovik-2012},
which arrived to the approximate value $\nu=1.5$ that is far from the exponent
obtained from numerical simulations of the respective models, $\nu=1.7$ and
$\nu=1.8$.

In this section we propose an approach that gives an insight into the
dynamics of the system towards consensus and provides a value of $\nu$ in good
agreement with simulations.  Equation~(\ref{tau}) shows that the mean
consensus time has a linear contribution ($\tau_1 \sim N$) that corresponds to
short-lived  realizations (type 1) and a non-linear term ($\tau_2 \sim
N^{1.71}$) corresponding to long-lived realizations (type 2).  Given that
$\tau_1$ is much smaller than $\tau_2$ for the explored range of $N$ (see
Fig.~\ref{tau-N}), we can assume that $\tau$ is mainly determined by the
long-lasting realizations that fall into a stripe state (type-2
realizations).  The evolution of a typical type-2 realization consists on two
different stages, as we can see from the evolution of $\rho$ in 
Fig.~\ref{rho-t}(b).  The initial stage is characterized by the dynamics of
domain coarsening where $\rho$ exhibits a power-law decay up to a time $t \simeq
10^4$.  Then, the system falls into a striped metastable state where $\rho$
stays nearly constant until consensus is reached at time 
$t \simeq 2 \times 10^6$.  Therefore, we see that the consensus time is 
greatly controlled by the duration of this stripe stage, given that it is much 
longer than the initial coarsening stage.  

\begin{figure}[t]
  \begin{center}
    \vspace{1cm}
    \begin{tabular}{cc}
      \includegraphics[width=5.7cm, bb=70 -20 550 550]{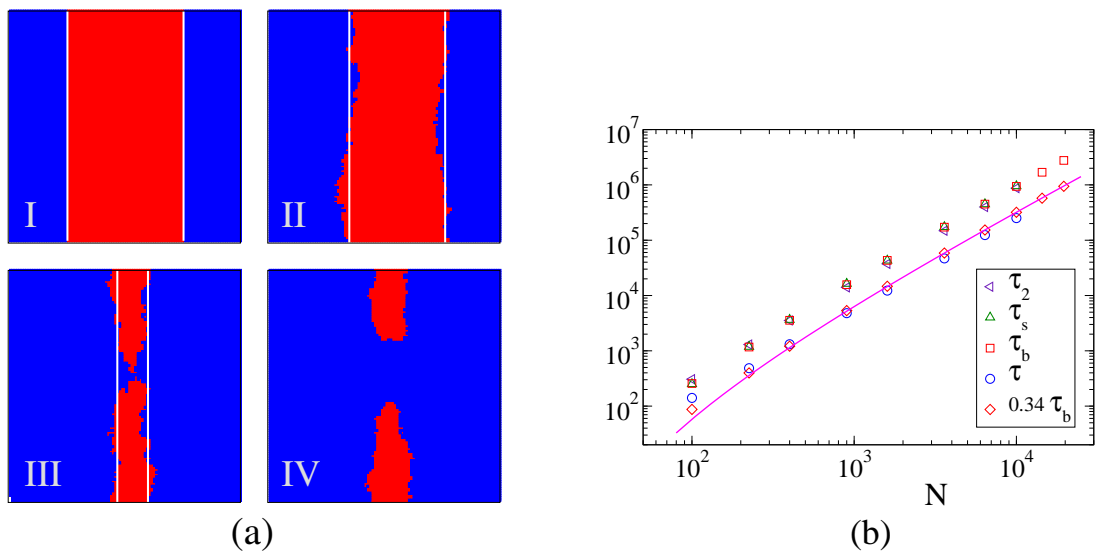} &
      \hspace{1.5cm}
      \includegraphics[width=4.7cm, bb=70 -20 550 550]{Fig7b.eps}
    \end{tabular}
    \caption{(a) Snapshots of a $100 \times 100$ square lattice at four different
      times, showing the evolution of same-opinion-orientation stripes in
      a single realization (negative opinions $-1$ and $-2$ in blue and
      positive opinions $1$ and $2$ in red).  Vertical straight lines denote
      the position of the stripe interfaces.  (b) We compare the mean consensus
      time of type-2 (stripe) realizations starting from a random initial
      condition $\tau_2$ (left triangles), with the mean consensus time
      $\tau_s$ (up triangles) and the mean interface breaking time $\tau_b$ 
      (squares) starting from the striped configuration showed in snapshot I
      of panel (a).  We also compare the mean consensus time $\tau$ (circles)
      with the estimation $0.34 \, \tau_b$ (diamonds).  The solid line is the 
      analytical approximation from Eq.~(\ref{tau4}).}     
    \label{stripe-tau}
  \end{center}
\end{figure}

To studied the dynamics of stripes 
we prepared the system in an initial condition that consisted on two vertical 
stripes of width $L/2$ each, as we see in Fig.~\ref{stripe-tau}(a-I). 
Figure~\ref{stripe-tau}(a) shows a typical evolution of the stripes in a single 
realization, where we combined
both opinions of a given orientation into a single color to make the interfaces look
more clear to the eyes ($-1$ and $-2$ in blue, $1$ and $2$ in red).  
The interfaces that separate the stripes freely diffuse in the direction
perpendicular to the interfaces [Fig.~\ref{stripe-tau}(a-II)] until they meet and
annihilate each other, cutting one stripe in two [Fig.~\ref{stripe-tau}(a-III)].
Then, during the last stage, the resulting domain quickly shrinks
[Fig.~\ref{stripe-tau}(a-IV)] and dissappear, and the system reaches consensus.
As this last stage is much shorter than the diffusive stage, the mean
consensus time starting from a stripe initial state, called $\tau_s$, can be
approximated as the mean time required for the two interfaces to meet and break, 
which we call the ``mean breaking
time'' $\tau_b$.  In Fig.~\ref{stripe-tau}(b) we verify that $\tau_b$ (squares)
is indeed very similar to $\tau_s$ (up triangles).  We also see that $\tau_b$
is similar to the mean consensus time $\tau_2$ of type-2 realizations starting
from a random initial condition (left triangles), as we suggested previously.
  Then, using Eq.~(\ref{tau}) we find that $\tau$ can be approximated as
\begin{equation} 
  \tau \simeq 0.34 \, \tau_b,
  \label{tau-taub}
\end{equation} 
represented by diamonds in Fig.~\ref{stripe-tau}(b).
Based on this result, we derive in subsection \ref{diffusion}
an analytical approximation for the dependence of $\tau_b$ with $L$ using the
diffusion properties of the interfaces, and in subsection \ref{roughness} we
improve this approximation by incorporating the roughness properties of the
interfaces.

\subsection{Estimation of $\tau_b$ considering two diffusive point-like particles}
\label{diffusion}

\begin{figure}[t] 
  \centerline{\includegraphics[width=12.0cm]{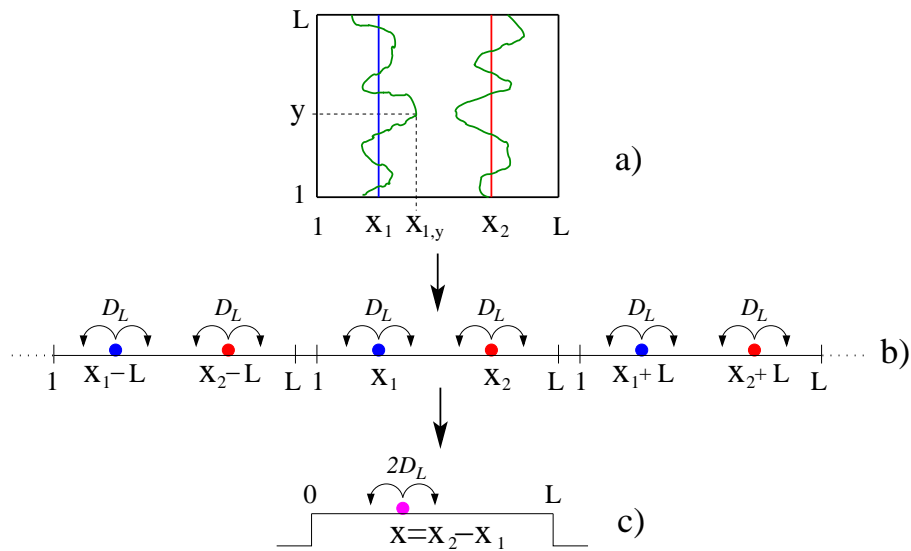}}
  \caption{Illustration of the mapping of the stripe interface dynamics to the 
    problem of two diffusive point-like particles in an interval $[1,L]$ with
    periodic boundary
    conditions.  a) Vertical lines indicate the positions $x_1$ and $x_1$ of
    the interfaces, denoted by circles (particles) in panel b).  b) The
    system is replicated in the entire $1D$ space, where the
    particles diffuse freely with no boundary constraints.  c) The equivalent
    particle with position $x=x_2-x_1$ diffuses in the interval $[0,L]$ with
    absorbing boundaries at the ends.  The diffusion coefficient $2 D_L$ is twice as
    that of particles in panel a).}
  \label{walkers}
\end{figure}

To study the dynamics of the interfaces we start by defining the position
$x_i(t)$ of interface $i$ ($i=1,2$) at a given time $t$ as the mean value of
the interface positions $x_{i,y}(t)$ at hight $y$ [see Fig.~\ref{walkers}(a)]
\begin{equation} 
  x_i(t) = \frac{1}{L} \sum_{y=1}^L x_{i,y}(t).
\end{equation} 
Then, we can interpret $x_1$ and $x_2$ as the respective positions of two
independent point-like particles that diffuse in an interval $[1,L]$ with 
periodic boundary conditions, which they annihilate when they meet.  
This equivalence was proposed by Chen and Redner in the MR
model \cite{Chen-2005}, and also used later by Volovik and Redner  in the
confident VM \cite{Volovik-2012}.  We have checked that particle 1 (and
also particle 2) moves diffusively by measuring the time evolution of the variance
of $x_1$,  $\sigma^2(t) = \langle x_1^2 \rangle(t) - \langle x_1 \rangle^2(t)$, where
averages were done over $10^4$ independent realizations.  We found that 
$\sigma^2(t)$ increases linearly with time for various linear sizes $L$ and that the
diffusion coefficient $D_L$, calculated from the relation 
$\sigma^2(t)=2 \, D_L \, t$ of a diffusive process, decays as $1/L$ (plots not
shown).  Indeed, we observed that all curves collapse when the $y$-axis is
rescaled by $L$, obtaining the approximate relation
\begin{equation} 
  D_L \simeq \frac{d}{L},
  \label{DL}
\end{equation} 
with $d=0.04$.  An estimation of this scaling relation was developed in 
\cite{Chen-2005,Volovik-2012} by assuming that each point at the interface 
$x_{i,y}$ behaves as an independent random walker
\cite{Plischke-1987,Spirin-2001-a} that jumps one site to the right or left with
equal probabilities.  Then, $\sqrt{D_L}$ should be proportional to the mean
displacement of the interface's position $ x_1$ in a time interval $\Delta
t=1$, which scales with the number of walkers $L$ as $\sqrt{L}/L=L^{-1/2}$,
thus $D_L \sim 1/L$.

\begin{figure} 
  \centerline{\includegraphics[width=7.5cm]{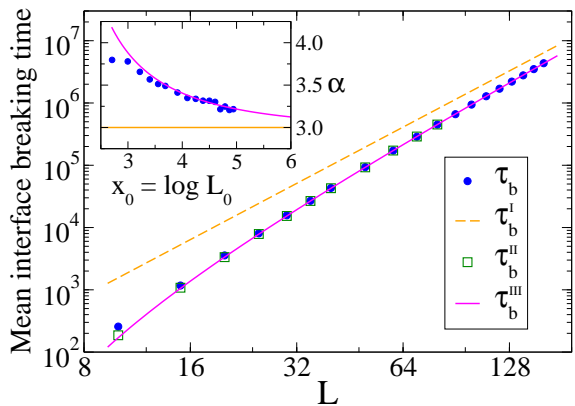}}
  \caption{Mean interface breaking time vs lattice side $L$ on a log-log scale
    starting from the stripe initial condition of Fig.~\ref{stripe-tau}(a-I).
    We compare simulation results ($\tau_b$, filled circles) with the following 
    approximate expressions: $\tau_b^{\mbox{\tiny I}}$ (dashed line) from 
    Eq.~(\ref{tau1}), $\tau_b^{\mbox{\tiny II}}$ (empty squares) from
    Eq.~(\ref{tau2}) and $\tau_b^{\mbox{\tiny III}}$ (solid line) from 
    Eq.~(\ref{tau3}).  Inset: Circles correspond to the local slope of the 
    $\tau_b$ vs $L$ curve on a log-log scale calculated from the data points of
    the main figure, while the solid line is the analytic approximation 
    Eq.~(\ref{slope}).}
  \label{tau-L}
\end{figure}

We can now approximate the mean interface breaking time $\tau_b$ as the mean
time the particles take to meet in the $[1,L]$ interval, when their
initial positions are a distance $L/2$ apart.  Given to the periodic
character of the interval's boundaries, it proves useful to consider an equivalent
system that is obtained by replicating the interval and particles in the
one-dimensional ($1D$) space [see Fig.~\ref{walkers}(b)], where particles can freely
diffuse in the entire $1D$ space without boundary constraints.  In this
replicated system, particle 1 moves always between particle 2-left and
particle 2 ($x_2-L \le x_1 \le x_2$) until it annihilates with one of these
two particles ($x_1= x_2-L$ or $x_1=x_2$).  Thus, the difference 
$x \equiv x_2- x_1$ can be seen as the position of an equivalent particle that
diffuses in the interval $[0,L]$ with absorbing boundaries at $x=0$ and $x=L$ 
[see Fig.~\ref{walkers}(c)].  Then, the problem is reduced to the escape of a 
particle with diffusion $2D_L$ (twice of that of particles 1 and 2) from an
interval $[0,L]$ starting from a position $x=L/2$, whose exact expression for
the mean exit time is known to be $L^2/16 D_L$ (see for instance 
\cite{Redner-2001}).  After replacing the expression Eq.~(\ref{DL}) for $D_L$
we obtain
\begin{equation}  
  \tau_b^{\mbox{\tiny I}} = \frac{L^3}{16 \, d},
  \label{tau1}
\end{equation} 
where the superindex $\mbox{I}$ in $\tau_b^{\mbox{\tiny I}}$ is
used to indicate a first-order approximation for  $\tau_b$ (see next subsection for
higher-order approximations).  In Fig.~\ref{tau-L} we compare the expression 
Eq.~(\ref{tau1}) for $\tau_b^{\mbox{\tiny I}}$ (dashed line) with the value
of $\tau_b$ obtained from numerical simulations (circles).  Even though we
see that $\tau_b^{\mbox{\tiny I}}$ is a reasonable approximation of $\tau_b$,
it overestimates $\tau_b$ for all simulated values of $L$.  However, we shall
show later that $\tau_b^{\mbox{\tiny I}}$ asymptotically approaches $\tau_b$
in the $L \to \infty$ limit.   This observation was already reported in
\cite{Chen-2005,Volovik-2012} together with the approximate scaling $\tau \sim
L^3 = N^{3/2}$.

\subsection{Estimation of $\tau_b$ considering two diffusive rod-like particles}
\label{roughness}

\begin{figure}[t]
\centerline{\includegraphics[width=15.5cm]{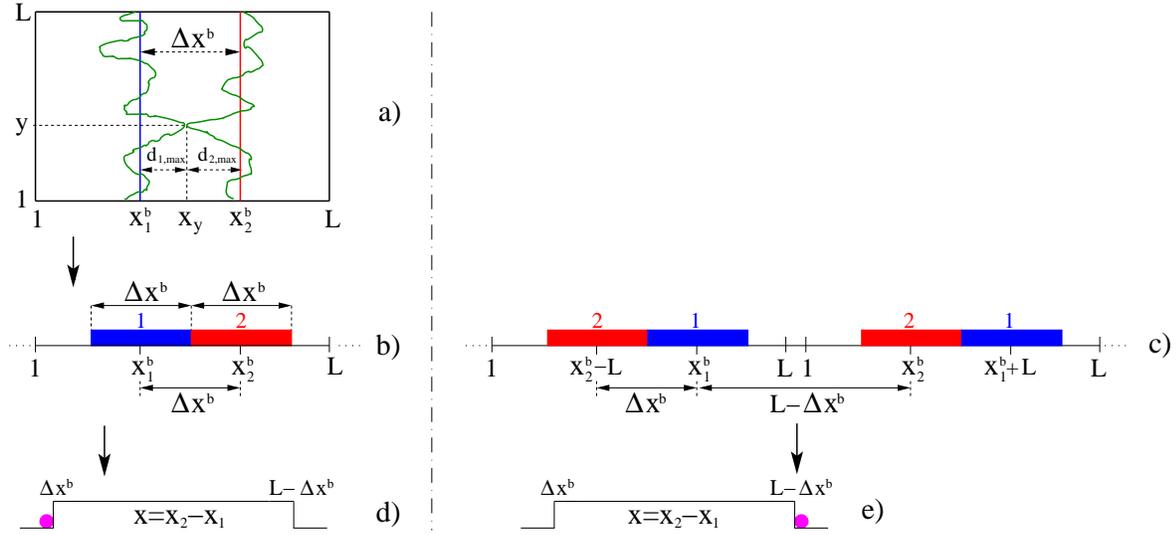}}
  \caption{a) The stripe interfaces break when they are a distance 
    $\Delta x^b$ apart and they touch at a point $y$ by the first time.  b)
    and c)  The interfaces are mapped to diffusive rod-like particles in the 
    replicated $1D$ space that annihilate each other when they collide.  d)
    and e)  The equivalent point-like particle with position $x=x_2-x_1$
    diffuses in the interval $[\Delta x^b,L-\Delta x^b]$ with absorbing
    boundaries at the ends.}
  \label{walkers-delta}
\end{figure}

The meeting time $\tau_b^{\mbox{\tiny I}}$ can be considered as a first
approximation  for $\tau_b$, where it is assumed that stripes' interfaces
break  when their positions become exactly the same ($x_1=x_2$).  However, this
approximation neglects the roughness of each interface, which plays an
important role in the breaking dynamics as we shall see.  A more refined
approximation that takes into account the width of the interfaces considers
that, in a given realization, the interfaces break when they are located at
some distance  $\Delta x^b=|x_2^b-x_1^b| > 0$ apart [see 
Fig.~\ref{walkers-delta}(a)], where  $x_1^b$ and $x_2^b$ are the respective
interfaces' positions at the breaking time.  The idea behind this argument is
that the breaking happens when the interfaces touch by the first time at some
point $y$ that depends on the specific roughness of the interfaces at that
moment, as we see in Fig.~\ref{walkers-delta}(a).  Therefore, each interface
can be better described by a diffusive rod-like particle of length $\Delta x^b$ that
represents the interfaces' width at the breaking moment 
[see Fig.~\ref{walkers-delta}(b)].
These two rods diffuse until they collide and annihilate in one of the two
possible ways shown in panels b) and c) of Fig.~\ref{walkers-delta}.  In the 
replicated system, the center of rod 1 moves between positions 
$x_1=x_2-\Delta x^b$ [panel b)] and $x_1=x_2-L+\Delta x^b$ [panel c)], and
thus the difference $x=x_2-x_1$ between the rods' centers describes the
position of a point-like particle that moves in the interval 
$[\Delta x^b,L-\Delta x^b]$ of reduced length $L-2\Delta x^b$ [panels (d) and (e)].

\begin{figure}[t]
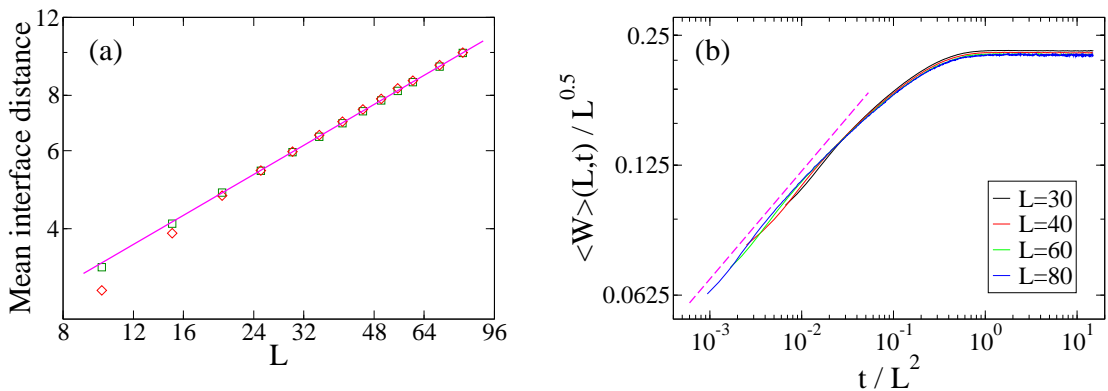

  \begin{center}
    \begin{tabular}{cc} \hspace{-3.0cm}
      \includegraphics[width=4.5cm, bb=70 -20 550 550]{Fig11a.eps} &
      \hspace{3.0cm}
      \includegraphics[width=4.5cm, bb=70 -20 550 550]{Fig11b.eps}
    \end{tabular}
    \caption{ a) Average distance between interfaces when they break
      $\langle \Delta x^b \rangle$ (squares) and average maximum interface deviation
      $2 \langle d_{max} \rangle$ (diamonds) as a function of $L$.  The solid
      line is the best power-law fit $L^{0.525}$ to $\langle \Delta x^b
      \rangle$.  (b) Growth of the average interface width $\langle W \rangle$ with
      time. The width and the time were rescaled by $L^{0.5}$ and $L^2$, 
      respectively, to obtain a data collapse for the linear sizes indicated
      in the legend.  The dashed line indicates the initial power law growth 
      $t^{0.25}$.}
    \label{Delta-width}
  \end{center}
\end{figure}

If we take the average value of $\Delta x^b$ over many realizations of the
dynamics, $\langle \Delta x^b \rangle$, as the effective distance between the
interfaces when they touch by the first time, the problem can be reduced to
the escape of a particle from an interval of ``effective length'' 
$\overline L=L-2 \langle \Delta x^b \rangle$.  Then, the mean escape time is 
$\overline L^2/16 \, D_L$ or, replacing the above expression for $\overline L$ and
Eq.~(\ref{DL}) for $D_L$, is
\begin{equation} 
  \tau_b^{\mbox{\tiny II}} = \frac{\left( 1- 2 \, \langle
        \Delta x^b \rangle L^{-1} \right)^2 L^3}{16 \, d}.
  \label{tau2}
\end{equation}  Equation~(\ref{tau2}) represents a second approximation that
incorporates the average distance between interfaces when they meet.  To test
Eq.~(\ref{tau2}) we run simulations and measured the average interface
distance $\langle \Delta x^b \rangle$ for several values of $L$ [squares in
Fig.~\ref{Delta-width}(a)].  The interface breaking moment of a given
realization  was taken as the time for which all sites of at least one lattice
row have either state $2$ or $-2$ by the first time.  Empty squares  in
Fig.~\ref{tau-L} represent the estimation $\tau_b^{\mbox{\tiny II}}$ of
$\tau_b$ obtained by plugging the numerical value of  $\langle \Delta x^b
\rangle$ into  Eq.~(\ref{tau2}), which is in good agreement with simulation
results (filled circles) for $L \ge 15$.  This shows that the roughness of the
interfaces plays a very important role in the breaking dynamics, leading to
large deviations of $\tau_b$ from the $L^3$ scaling law (dashed line in
Fig.~\ref{tau-L}) as $L$ decreases.  These deviations, which become very
visible for low $L$, are captured rather well by the prefactor   $\left( 1-2
\,\langle \Delta x^b \rangle/L\right)^2$ of $\tau_b^{\mbox{\tiny II}}$ in
Eq.~(\ref{tau2}).   We see in Fig.~\ref{Delta-width}(a) that $\langle \Delta
x^b \rangle$ grows with $L$ as $L^{0.525}$ (solid line), and thus the ratio
$\langle \Delta x^b \rangle/L$ vanishes as $L$ increases,  leading to the
expression Eq.~(\ref{tau1}) for $\tau_b^{\mbox{\tiny I}} $ and confirming the
hypothesis that Eq.~(\ref{tau1}) is correct in the $L \to \infty$ limit.   As
we show in \ref{Deltax}, the exponent $0.525$ is related to the
\emph{roughness exponent} $\alpha \simeq 0.5$ associated to the saturation
value of the interfaces' width.   An interesting insight from
Eq.~(\ref{tau2}) is that a pure power law $\tau_b^{\mbox{\tiny II}} \sim L^{2
\nu}=N^{\nu}$ is never obtained for a finite  value of $L$.  Instead, the
correction factor   $\left( 1-2 \,\langle \Delta x^b \rangle/L\right)^2$
introduces a downward curvature in the $\tau_b^{\mbox{\tiny II}}$ vs $L$ curve
on a double logarithmic scale, which decreases with $L$ and becomes very small
for $L \gtrsim 40$ (see Fig.~\ref{tau-L}).  As a result, the data can be well
fitted by a power law function of $N$ with an effective exponent $\nu > 1.5$,
as those shown in Fig.~\ref{tau-N} for $\tau$ and $\tau_2$.

By plugging the power-law approximation  $\langle \Delta x^b \rangle \simeq
L^{0.525}$ into Eq.~(\ref{tau2}) we obtain the following approximate
expression for the mean breaking time:
\begin{equation} 
  \tau_{b}^{\mbox {\tiny III}} = \frac{\left( 1- 2 L^{-0.475} \right)^2 L^3}{16 \, d}.
  \label{tau3}
\end{equation} 
As we can see in Fig.~\ref{tau-L}, Eq.~(\ref{tau3}) represented
by a solid line fits the numerical data (filled circles) very well for  $L
\gtrsim 15$.  Finally, using Eq.~(\ref{tau-taub}) we arrive to the
approximate expression
\begin{equation} 
  \tau \simeq \frac{0.34 \, \left( 1- 2 N^{-0.2375} \right)^2
    N^{1.5}}{16 \, d}.
  \label{tau4}
\end{equation} 
for the mean consensus time.  Equation~(\ref{tau4}) is plotted
by a solid line in Fig.~\ref{stripe-tau}(b).  We see that, even though there are
some discrepancies with numerical results (circles), Eq.~(\ref{tau4}) captures
rather well the behavior of $\tau$ with the system size for almost the entire
range of $N$ values.

We can now exploit the approximate functional form of $\tau_b$ given by
Eq.~(\ref{tau3}) to analyze the scaling of $\tau_b$ for a wide range of $L$.
The factor $L^{-0.475}$ introduces a downward curvature in
$\tau_b^{\mbox{\tiny III}}$ --when plotted in log-log scale-- that vanishes as
$L$ increases.  Therefore, we can approximate the shape of
$\tau_b^{\mbox{\tiny III}}$ around a given value $L_0$ as a power law of $L$ (see
\ref{power-law} for calculation details)
\begin{equation} 
  \tau_b^{\mbox{\tiny III}}(L,L_0) \simeq A(L_0) \, L^{\alpha(L_0)},
  \label{tau3-2}
\end{equation}  
where 
\begin{eqnarray} 
  \label{A}
  A(L_0) &=& \frac{1}{16 \, d} \left(1-2 \,L_0^{-0.475}
  \right)^2  L_0^{\frac{1.9}{\left(2-L_0^{0.475} \right)}} ~~~~ \mbox{and} \\
  \label{slope} 
  \alpha(L_0) &=& 3 + \frac{1.9}{\left(L_0^{0.475} -2 \right)}.
\end{eqnarray} 
We can check that in the thermodynamic limit $L_0 \to \infty$
the exponent $\alpha(L_0)$ approaches the value $3.0$ as previously suggested,
while $A(L_0)$ approaches $1/(16\,d)$, recovering the approximation
$\tau_b^{\mbox{\tiny I}} \simeq L^3/(16\,d)$ from Eq.~(\ref{tau1}).  The
exponent $\alpha(L_0)$ from Eq.~(\ref{slope}), which measures the slope of the 
$\log[\tau_b^{\mbox{\tiny III}}(L)]$ vs $\log(L)$ curve at some point $\log(L_0)$, is
plotted by a solid line in the inset of Fig.~(\ref{tau-L}) and compared to the
numerical value (filled circles) obtained by calculating the local slope of
the $\tau_b$ data points from the main figure.  We can see that the slope
decreases very slowly with
$L_0$, and thus for the values of $L$ measured in simulations $\alpha$ stays
nearly constant and can be approximated by a clean power law.  Then, we can
use Eq.~(\ref{slope}) to approximate the mean breaking time as  $\tau_b \sim
L^{\alpha}=N^{\alpha/2}$ in the range of system sizes used in simulations, and
compare $\alpha$ with the numerical exponents obtained from Fig.~\ref{tau-N}
by fitting
the numerical data with a power law.  For instance, the slope at $N=2000$
($L \simeq 45$) from Eq.~(\ref{slope}) is $\alpha/2 \simeq 1.73$, which agrees
quite well with the numerical slope $1.71$ for $\tau_2$ in the range 
$4 \times 10^2 \le N \le 10^4$.  The theoretical value $\alpha/2$ is also a fair
approximation of the numerical exponent $1.64$ obtained from the $\tau$ vs $N$
data (only $5.5\%$ off), even though we expect that this approximation improves
for larger values of $N$.  Finally, we also note that it turns very difficult
to reach a slope close to $1.5$ in simulations of the model,
because of the very slow decrease of $\alpha$ with $L_0$.  For instance, to
achieve a slope smaller than $1.545$ (less than $3\%$ difference with $1.5$)
Eq.~(\ref{slope}) predicts that we would need to run simulations in systems
with linear dimension $L \gtrsim 750$, whose consensus times are of order
$\tau \sim 10^{9}$ (Eq.~\ref{tau3}), which is almost impossible to achieve in
reasonable computation times.

\section{Summary and conclusions}
\label{summary}

We studied an agent-based model on a $2D$ lattice that explores the
competition between persuasion and compromise in opinion formation.  We found
that nearest-neighbor interactions between agents induce a very rich  domain
coarsening dynamics, which plays a fundamental role in the evolution of the
system and the approach to consensus.  The properties of the coarsening
strongly depend on the relative frequency  between persuasion and compromise
events, measured by the ratio $r=p/q$ between persuasion and compromise
interaction probabilities.    When the compromise process dominates over the
persuasion process the dynamics is akin to that of the VM during an
initial short transient, in which domains are formed by moderate agents and
the coarsening is without surface tension.  This is associated to a
centralized opinion state where most agents adopt moderate opinion values.
Domain growth eventually leads to a state where all agents have the same
opinion orientation (positive or negative).  Then, moderate agents start to
become extremists and the system displays a slow  exponential approach to
consensus in an extreme opinion that is achieved in a time that scales as
$r^{-1} \ln N$ with the population size $N$.  In the opposite case scenario where
persuasion dominates over compromise, the coarsening is driven by surface
tension and moderate agents are located at the interface between  domains
formed by extremists.  This corresponds to a polarized opinion state in which
the population is divided into two groups that adopt extreme and opposite
opinions (positive and negative).  The final approach to consensus can be very
long if the system falls into a striped metastable configuration, where the two
interfaces that define a stripe diffuse until they meet and annihilate.  The
mean consensus time of this type of realizations scales as  $N^{1.71}$.  When
the average is done over all realizations, which include short-lived
realizations with a lifetime that scales as $N$, the scaling of the overall
mean consensus time is $\tau \sim N^{\nu}$, with $\nu=1.64$.

An insight into the approach towards consensus of striped configurations was
obtained by mapping the dynamics of stripes into the problem of two rods that
freely diffuse in $1D$ and annihilate when they collide by the first time.
This method takes into account the width of stripe interfaces, which becomes
relevant when interfaces meet and break.  An analytical estimation of the mean 
collision time using known results on first-passage problems allowed to obtain
the approximate expression Eq.~(\ref{tau3}) for the mean lifetime of stripes, which
is in good agreement with results from simulations of the model.  
Also, Eq.~(\ref{tau4}) for the mean consensus time shows that the scaling 
$\tau \sim N^{\nu}$ is an
approximation obtained by fitting with a power-law the numerical data over a
finite range of $N$, given that the effective exponent $\nu$ around a given
$N$ decreases and approaches the value $1.5$ in the thermodynamic limit 
($N \to \infty$).  These results show that analytical deviations from the
scaling exponent $\nu=1.5$, obtained by assuming that interfaces behave as 
point-like particles, are due to the roughness of the interfaces.

In summary, the $2D$ spatial topology of interactions has a large impact on
the behavior of the M-model respect to the MF case.  Opinion bi-polarization is
much more stable in lattices than in MF, due to the existence of long-lived
metastable states with a spatial pattern of opinions that consists on two
stripes composed by both types of extremists.  This dynamics leads to consensus
times in lattices that are much longer than those obtained in a MF setup.
The width of the interfaces between stripe domains plays an important role in
the dynamics close to consensus, when interfaces are about to annihilate each
other.  Taking into account the scaling properties of the interface width
allows to derive an expression for the behavior of the
mean consensus time with the system size, in good agreement with simulations.
This expression provides an explanation for
the non-trivial numerical exponent $\nu=1.64$, and also for similar exponents 
observed in related models where consensus is reached by curvature driven 
coarsening.  Another observation is that the bi-polarization is found for $p>q$
in MF, while in lattices is found for much lower values of persuasion,
approximately for $p>q/3$.  Therefore, a small level of homophily is enough to
induce bi-polarization in a population that interacts in lattices.  Thus, the
lattice topology seem to intensify the effect of homophily and PAT on the
emergence of bi-polarization.  This result resembles that obtained in the
Scheling model for racial segregation \cite{Schelling-1971}, where even a
small preference to have neighbors of the same race on a lattice is able to
induce a large spatial  segregation of the population into same-race domains.

It would be worthwhile to study the dynamics of the M-model on complex
networks of different kinds, which are more realistic descriptions of the
topology of social interactions among people.  It would also be interesting to
investigate the role of the network connectivity in the propagation and
ultimate dominance of an extreme opinion \cite{Vazquez-2010}.  Finally, a
natural extension of the model would include variations of the persuasion
and/or compromise rules that could enhance bi-polarization in lattices or in
general topologies.

\section*{Acknowledgments}
We would like to thank Gabriel Baglietto for helpful discussions.  We
also acknowledge financial support from CONICET (PIP 0443/2014).

\clearpage
\appendix

\section{Analysis of the mean interface breaking distance  $\langle
\Delta x^b \rangle$}
\label{Deltax}

We can gain an insight into the scaling 
$\langle \Delta x^b \rangle \simeq L^{0.525}$ by relating the distance between 
interfaces at the breaking
moment with the properties of the interface roughness, as we illustrate in
Fig.~\ref{walkers-delta}(a).  More precisely, the interfaces touch at a hight
$y$ ($x_{1,y}=x_{2,y}=x_y$) where the respective interface deviations from
$x_1$ and $x_2$ reach their maximum values $d_{1,max}=|x_y-x_1^b|$ and 
$d_{2,max}=|x_y-x_2^b|$.  Therefore, the distance between interfaces can be 
approximated as  $\Delta x^b \simeq d_{1,max}+d_{2,max}$.  Given that the high
$y$ of the touching point
varies among realizations, we calculated the average value of the maximum
deviation  $\langle d_{max} \rangle$ at both sides of each interface over many
realizations of the dynamics.  Results are shown in Fig.~\ref{Delta-width}(a)
(diamonds).  We see that $2 \, \langle d_{max} \rangle$ agrees very well with
$\langle \Delta x^b \rangle$ for $L \gtrsim 20$, and that follows the
power-law scaling $\langle d_{max} \rangle \sim L^{0.525}$ (solid line).  We
speculate that this scaling is related to the scaling properties of the width
of the interfaces, defined as the standard deviation of the
interface positions $x_{i,y}$ along the y-axis \cite{Family-1986,Barabasi-1995}
\begin{equation} 
  W_i = \left[\frac{1}{L} \sum_{y=1}^L x_{i,y}^2 - \left(
      \frac{1}{L}  \sum_{y=1}^L x_{i,y} \right)^2 \right]^{1/2} =  \left[\frac{1}{L}
    \sum_{y=1}^L (x_{i,y}-x_i)^2  \right]^{1/2}.
\end{equation} 
The time evolution of the average interface width calculate over many
realizations $\langle W \rangle$ [see Fig.~\ref{Delta-width}(b)] has an
initial stage in which  $\langle W \rangle$ grows as $t^{\beta}$, followed by
a second stage where $\langle W \rangle$ reaches a saturation value (plateau)
that increases with $L$ as
$\langle W \rangle_{\mbox{\tiny sat}}(L) \sim L^{\alpha}$, where $\alpha
\simeq 0.5$ is the \emph{roughness exponent} and  $\beta \simeq 0.25$ is the
\emph{growth exponent}.  These exponents are consistent with those of the
Edwards-Wilkinson universality class of surface growth \cite{Barabasi-1995}.
Indeed, by an appropriate rescaling of the $x$ and $y$ axis the data can be
collapsed into a single function [Fig.~\ref{Delta-width}(b)] showing that the
interface growth obeys the Family-Vicsek scaling relation $\langle W
\rangle(L,t) =  L^{\alpha} f(t/L^z)$ with $z=\alpha/\beta \simeq 2$, and
$f(x) \sim x^{\beta}$ for $x \ll 1$ and $f(x) = \mbox{constant} \simeq 0.22$
for $x \gg 1$.  We note that the same scaling behavior of the interface
dynamics was reported in \cite{DallAsta-2008} for a broad family of voter
models with intermediate states, as the present M-model.  We have checked that
the width has already reached its saturation value at the mean breaking time
$\tau_b$, given that $\tau_b$ is much longer than the ``crossover time'' that
separates the growth and the saturation stages.  Therefore, one expects that
the maximum deviation of the interface should be proportional to the
saturation value of the interface width, leading to the approximate scaling 
$\langle d_{max} \rangle \sim L^{0.5}$.  We do not know how to explain the
small discrepancy with the scaling $\langle d_{max} \rangle \sim L^{0.525}$
obtained from simulations.

\section{Approximation of $\tau_b^{\mbox{\tiny III}}$ as a power law}
\label{power-law}

To obtain the coefficient $A(L_0)$ and the exponent $\alpha(L_0)$ of the
power-law approximation 
\begin{equation} 
  \tau_b^{\mbox{\tiny III}}(L,L_0) \simeq A(L_0) \, L^{\alpha(L_0)}
  \label{tau3-22}
\end{equation}  
of $\tau_b^{\mbox{\tiny III}}$ from Eq.~(\ref{tau3}) it proves useful to
work on a double logarithmic scale, where Eq.~(\ref{tau3-22}) becomes the
straight line
\begin{equation} 
  y(x,x_0) \simeq \log[A(x_0)] + \alpha(x_0) \, x
  \label{yx}
\end{equation} 
in the variable $x \equiv \log(L)$, with $x_0 \equiv \log(L_0)$
and  $y(x,x_0) \equiv \log[\tau_b^{\mbox{\tiny III}}(L,L_0)]$.  Then,
rewriting Eq.~(\ref{tau3}) in terms of the variables $x$ and $y(x)$ 
\begin{eqnarray*} 
  y(x)&=&2 \log \left(1-2 e^{-0.475\,x} \right) + 3\,x - \log(16\,d), 
  \label{yx-1}
\end{eqnarray*} 
and Taylor expanding $y(x)$ to first order in $x-x_0$ we obtain
\begin{eqnarray} 
  y(x)&\simeq& 2 \log \left(1-2 e^{-0.475\,x_0} \right)+
  \frac{1.9\,(x-x_0)}{\left(e^{0.475\,x_0} -2 \right)} + 3 x - \log(16\,d)
  \\
  &=& \log \left[ \frac{\left(1-2 e^{-0.475\,x_0} \right)^2}{16 \, d} \right]- 
  \frac{1.9\, x_0}{\left(e^{0.475\,x_0} -2 \right)} +  
  \left[ 3 + \frac{1.9}{\left(e^{0.475\,x_0} -2 \right)} \right] x. \nonumber 
  \label{yx-2}
\end{eqnarray} 
Matching the coefficients of Eq.~(\ref{yx-1}) with those of  
Eq.~(\ref{yx}) and transforming back to the variable $L_0=e^{x_0}$ we arrive
to the expressions for $A(L_0)$ and $\alpha(L_0)$ quoted in Eqs.~(\ref{A}) and 
(\ref{slope}), respectively, of the main text.

\section*{References}

\bibliographystyle{unsrt}

\bibliography{references}

\end{document}